# Entangled-Photon Coincidence Fluorescence Imaging

**Giuliano Scarcelli and Seok H. Yun**

[1] *Harvard Medical School and Wellman Center for Photomedicine, Massachusetts General Hospital, 50 Blossom St., Boston, MA 02114, USA*

**ABSTRACT**

We describe fluorescence imaging using the second-order correlation of entangled photon pairs. The proposed method uses the fact that one photon of the pair carries information on where the other photon has been absorbed and has produced fluorescence in a sample. Because fluorescent molecules serve as "detectors" breaking the entanglement, multiply-scattered fluorescence photons within the sample do not cause image blur. We discuss experimental implementations.

**INTRODUCTION**

Entangled photons have recently attracted considerable interest in the optics community for potential applications in precision metrology, information processing, and imaging [1,2]. Coincidence imaging, or ghost imaging, was pioneered ten years ago, using entangled photon pairs that are typically generated via spontaneous parametric down conversion (SPDC) [3,4]. In this technique, one photon of the pair probes a distant object, while the other serves as a reference. The correlation in position and momentum of the entangled photons allows the object to be imaged remotely by measuring the position of the reference photon coincident with the probe photon. Classical light sources also can be used for coincidence imaging even though they cannot replicate all the features the entangled source can produce [5-7]. To date, coincidence imaging has been studied with simple intensity masks or phase objects [8].

In this paper, we propose a new method to extend the entangled coincidence imaging to fluorescent samples. We describe the principle and implementations of the method and compare it with classical imaging techniques widely used in biological applications.

**WORKING PRINCIPLE**

Consider the imaging setup shown in Fig. 1(a). A pair of entangled photons is generated by SPDC. The probe photon is directed to a sample containing fluorescence molecules. One of the molecules may absorb the probe photon, be excited, and generate a new photon with the Stokes frequency (wavelength) shift. This fluorescence photon is then detected by a "bucket" photon-counting detector (D1). A dichroic filter allows only fluorescence photons to be registered, rejecting any probe photons that have passed through or scattered off the sample without being absorbed. On the other hand, the reference photon is detected by an array of photon-counting detectors, and the position and time of arrival are registered. Because the entangled photons are simultaneously generated in the SPDC crystal, an ideal lossless system ensures that whenever D1 detects a fluorescence photon, one of the detector elements in D2 will capture the reference photon (the reverse would not be true if the probe-to-fluorescence conversion did not occur). The position and momentum of the reference and probe photons are correlated to each other. Therefore, from the measured location of the coincident reference photon, one can obtain information on where the probe photon was absorbed, i.e. the location of the fluorophore. By repeating such a measurement for many entangled photon pairs generated sequentially, one can produce a fluorescence image of the sample.

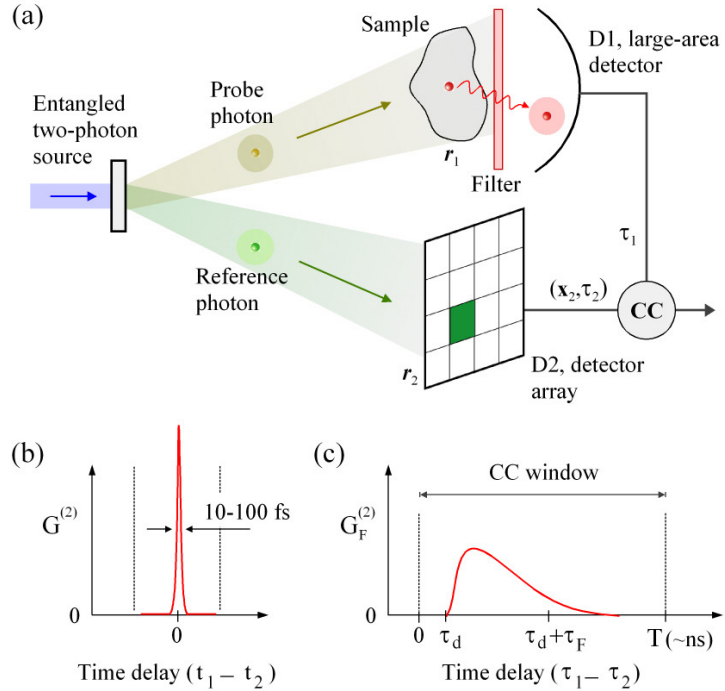

Fig. 1. Principle of entangled-photon fluorescence imaging. (a) Imaging system schematic. (b) The intrinsic correlation function of entangled photon pair measured with ideal fast detectors. (c) The correlation function broadened by fluorescence molecules and detectors. $\tau_F$ and $\tau_d$ are time delays due to the fluorescence lifetime and the photon propagation from the molecule to the detector. The integration time T is chosen to be longer than these time constants.

The exact time difference between detecting the reference and fluorescence photons, $\tau_1 - \tau_1$, is somewhat uncertain because of random optical and electrical time delays in the system. For example, fluorescence emission is a stochastic process with a lifetime, and the photon-counting module has a finite electronic timing jitter. These timing uncertainties, however, can be accommodated by having a correlation time window, $T$, longer than these time delays [Fig. 1 (b) and (c))]. In each correlation window, no more than one entangled photon pair should be dealt with, ideally in order to avoid any erroneous correlation count.

## THEORY

Applying Glauber's quantum theory of photodetection [9], we point out that the photo-absorption by fluorescence molecules must be regarded as the "detection" event upon which the probe photon is annihilated, the quantum state of the entangled photon pair is determined, and any correlation or coherence between the probe and reference photons is terminated. Any processes after the absorption, including fluorescence emission, the propagation of fluorescence photon, and photoelectric conversion at D1, merely serve as couriers delivering the detected signal with time delay.

Using Glauber's mathematical formalism of the second-order correlation, we express the coincidence rate function as:

$$I^{(2)}(\vec{r}_1;\vec{r}_2) = \iint_T d\tau_1 d\tau_2 \, G_F^{(2)}(\vec{r}_1,\tau_1;\vec{r}_2,\tau_2)/T, \tag{1}$$

where

$$G_F^{(2)} = \iiint dt_1 dt_2 dt_F \, G^{(2)}(\vec{r}_1,t_1;\vec{r}_2,t_2) F(\vec{r}_1,t_F-t_1) \gamma_1(\tau_1-t_F) \gamma_2(\tau_2-t_2). \tag{2}$$

Here $G^{(2)}$ is the intrinsic second-order correlation function computed in the case of ideal detectors, whereas $G_F^{(2)}$ is the correlation function taking into account the temporal convolution imposed by the finite response time of fluorescence generation and photoelectric conversion; $F(\vec{r},t)$ denotes the probability function of fluorescence generation, $F(\vec{r},t) = f(\vec{r})\exp(-t/\tau_F)H(t)/\tau_F$ where $f(\vec{r})$ describes the spatial distribution and concentration of fluorescence molecules, $\tau_F$ is the fluorescence lifetime, typically 0.2 – 5 ns, and $H(t)$ is the Heaviside step function. $\gamma_{1,2}(t)$ denotes the response function of the photo-electric detectors, that can be approximated to a Gaussian function with a width of 0.2 – 1 ns. D1 integrates all the fluorescence photons emitted in a sample volume *V*. Therefore, the final measured quantity is the marginal coincidence counting rate $R_C$:

$$R_C(\vec{r}_2) = \int_V d\vec{r}_1 \, I^{(2)}(\vec{r}_1;\vec{r}_2) \tag{3}$$

The purpose of the setup is to retrieve $f(\vec{r}_1)$ from the measurement of $R_C(\vec{r}_2)$.

Figure 2 shows an equivalent ray-optic diagram [4] of an imaging setup, similar to Fig. 1(a), in which an objective lens is inserted in the probe arm. In a simplified picture, the probe and reference photons are generated at the same time and at the same location $\vec{r}_c$ in the SPDC crystal and have perfectly phase-matched wave vectors, $\vec{k}_1$ and $\vec{k}_2$, i.e. $\vec{k}_1 + \vec{k}_2 = \vec{k}_P$ where $\vec{k}_p$ is the wave vector of the pump photon. From these properties, the beam propagation paths linking two points $\vec{r}_1$ and $\vec{r}_2$ can be identified, as represented by a red shaded region in Fig. 2. From the quantum optical viewpoint, the second-order correlation is established as a coherent linear superposition of all of the correlation beam paths. Nevertheless, the classical ray picture is adequate to understand the imaging relationship as follows. The phase matching condition leads to a Snell's-law-like relation at the crystal interface: $\lambda_2 \sin\theta_1 = \lambda_1 \sin\theta_2$ where $\lambda_1$ and $\lambda_2$ denote the wavelengths of probe and reference photons, respectively. Simple ray tracing leads to an imaging equation [10]:

$$\frac{1}{s_1} + \frac{1}{a+(\lambda_2/\lambda_1)b} = \frac{1}{f_{obj}}, \qquad (4)$$

where $f_{obj}$ denotes the focal length of the objective lens. This relation asserts that the imaging plane distance, $s_1$, can be varied in the sample remotely by changing the distance, $b$, of the reference detector. The wavelength dependence in Eq. (4) plays a similar role as chromatic aberration and may cause image blur if broadband spectra are used. This problem, however, can be avoided by spectral selection of correlated pairs or with an infinite conjugate imaging setup that employs an imaging lens in front of D2. On the other hand, the wavelength dependence may be used advantageously in conjunction with wavelength tuning to allow the imaging focal plane to be adjusted, even without moving the reference detector.

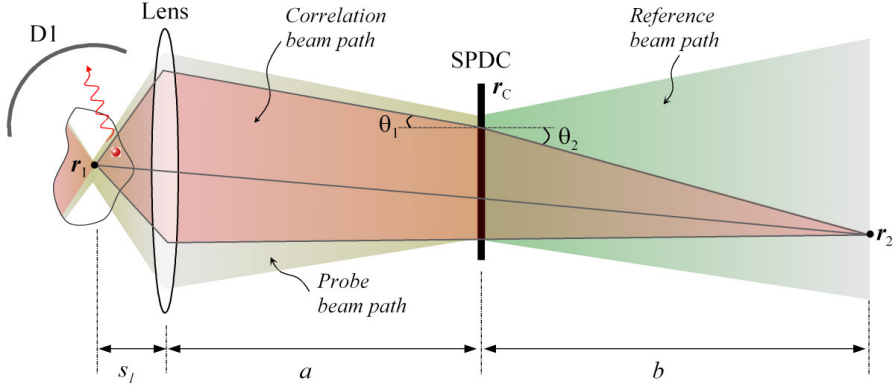

Fig. 2. Unfolded geometrical beam paths. The brown and green shaded regions depict all the possible optical paths of the probe and reference photons. Solid line traces (black) represent correlation beam paths between $r_1$ and $r_2$.

Any fluorescence molecules within the beam correlation region (red shade) can be excited by the probe photon with probability proportional to the local optical intensity, but any fluorescence photon emitted from this region, regardless of its true origin, will be assigned to $\vec{r}_1$, the conjugate location to $\vec{r}_2$. In this sense, the imaging system considered here is similar to classical wide-field fluorescence imaging.

Using Fourier optics formalism [10], the coincidence rate function can be expressed as

$$I^{(2)}(\vec{r}_1;\vec{r}_2) \approx f(\vec{r}_1)\left|\int d\vec{r}_c h_1(\vec{r}_1;\vec{r}_c) h_2(\vec{r}_2;\vec{r}_c)\right|^2, \qquad (5)$$

where $h_{1,2}$ are the 3D classical impulse response function of the probe and reference arms, respectively. When the imaging relationship given by Eq. (4) is established with 1:1 magnification, we get $I^{(2)}(\vec{r}_1;\vec{r}_2) \approx f(\vec{r}_1)\delta(\vec{r}_1-\vec{r}_2)$ and $R_C(\vec{r}_2) \propto f(\vec{r}_2)$ from Eq. (3). Imaging resolution can be analyzed by considering a point object $f(\vec{r}_1) = \delta(\vec{r}_1)$. From standard diffraction calculation, it can be shown that the system's resolution is identical to that of a classical standard microscope using the wavelength of $\lambda_1$.

**PRACTICAL CONSIDERATIONS**

Considering fluorescence dye solution as a sample, we estimate the maximum achievable coincidence counting rate to be $R_C \approx N_0(1-e^{-m\varepsilon L})\eta_F\eta_c\eta_1\eta_2$, where $N_0$ is the number of entangled photon pairs generated per second, $m$ the molar concentration, $\varepsilon$ the extinction coefficient, $L$ the sample thickness, $\eta_F$ the quantum efficiency of the fluorophore, $\eta_c$ the geometrical collection efficiency of the bucket detector D1, and $\eta_{1,2}$ are the quantum efficiency of the detectors. As an example, we consider a near-infrared fluorescence dye, Alexa-Fluor 700 (Invitrogen), which is widely used in biological imaging. Its peak absorption wavelength 700 nm is suited for a SPDC source pumped by an Argon laser at 351 nm [3]. The dye has $\varepsilon = \sim20$ M$^{-1}\mu$m$^{-1}$, $\eta_F = 0.25$, and $\tau_F = 1$ ns. Let us consider an experiment where D2 is a large scale array [10]. In practice, we can have $N_0 = 4\times10^6$ pairs per second, $T = 10$ ns, and a pixel dead-time of 50 ns. Based on the Poisson distribution, the error probability of having more than one photon pair in 10 ns is only 0.08 %. For a sample with $m = 100$ μM and $L = 20$ μm, we estimate that about 4% of probe photons are absorbed in the sample. Using $\eta_c = 0.5$, and $\eta_{1,2} = 0.7$, we get $R_C = 10,000$ counts per second. At this rate, the pixel acquisition time for collecting average 100 counts per pixel would be ~10 ms. Therefore, it will take about 100 sec to acquire an 8-bit 100x100 image. Alternatively, a single- or few-element detector can be scanned in 2D at the expense of acquisition time. Classical light sources, such as thermal chaotic light [6] and synchronized momentum-correlated sources [5], could be used for coincidence fluorescence imaging. However, only quantum correlations ensure both background-free coincidence and remote tunability of the imaging plane distance, $s_1$ [7].

**SPECTRALLY-ENCODED FLUORESCENCE SPECTROSCOPY AND IMAGING**

Next, we briefly describe the possibility of using entanglement in frequency for fluorescence measurement. Phase matching in SPDC leads to $\omega_1 + \omega_2 = \omega_P$ where $\omega_{1,2,P}$ are the frequencies of probe, reference, and pump photons, respectively. This indicates that one can determine $\omega_1$, without directly analyzing the spectrum of the probe photon, by measuring $\omega_2$ of the reference photon [12]. Combined with the coincidence counting measurement, this principle can be used for fluorescence absorption spectroscopy and imaging (Fig. 3) based on spectral encoding [13].

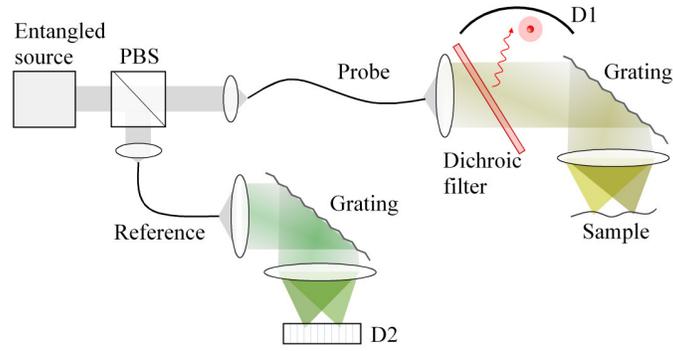

Fig. 3. Spectrally-encoded coincidence imaging. Orthogonally polarized entangled photons are separated into a fiber-optic probe and reference arms. The probe photon is spectrally dispersed and focused to a sample. Fluorescence photon (red) is collected by a large-area bucket detector, D1, and the position of the excited fluorescence molecule is determined by measuring the frequency (wavelength) of reference photon.

**DISCUSSION AND CONCLUSIONS**

Coincidence imaging offers a number of interesting features. First, our method is based on the second-order correlation measurement. To date, only direct detection (wide-field, confocal, or two-photon microscopy) and the first-order correlation ($4\pi$ microscopy) have been used for fluorescence imaging. Second, only a single-element detector is required in the probe arm. This may be advantageous where a 2D detector array is not readily available at the fluorescence wavelength or when it is difficult to place it near the sample; for instance, in endoscopy. Third, multiple scattering of fluorescence photons in a sample does not blur images. Figure 4(a) illustrates a situation in which a fluorescence photon undergoes elastic scattering before reaching the detector D1. Scattering of the fluorescence photon may affect the detection time $\tau_1$, but does not perturb the image retrieval because fluorescence is an incoherent process so that no position correlation exists between the fluorescence and reference photons. On the other hand, any elastic scattering events of the probe photon before it reaches the fluorophore (Fig. 4(b)) does affect the correlation property between the probe and reference photons, resulting in speckle-type image degradation. Two-photon microscopy also uses a bucket detector to collect fluorescence photons generated by nonlinear absorption, but it requires a high numerical aperture (NA) objective lens, mode-locked femtosecond laser, and beam scanning device [14].

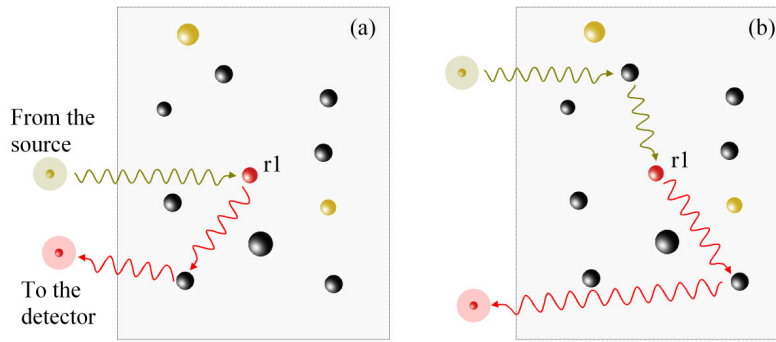

Fig. 4. The effect of multiple scattering in a sample.

In conclusion, we have proposed, for the first time to our knowledge, a method for fluorescence imaging and spectroscopy based on coincidence measurement of entangled photons. While our discussion here focuses on fluorescence, the principle can be applied to probing inelastic scattering, such as Raman scattering.